\documentstyle[12pt,preprint]{aastex}

\begin{document}
\title{The gas temperature in the surface layers of protoplanetary disks}
\author{I.~Kamp}
\affil{Space Telescope Science Institute, Baltimore, MD 21218; e--mail: 
kamp@stsci.edu}

\and

\author{C.P.~Dullemond}
\affil{Max Planck Institut f\"ur Astrophysik, P.O.~Box 1317, D--85741 
Garching, Germany; e--mail: dullemon@mpa-garching.mpg.de}


\begin{abstract}
Models for the structure of protoplanetary disks have so far been
based on the assumption that the gas and the dust temperature are equal. The
gas temperature, an essential ingredient in the equations of hydrostatic
equilibrium of the disk, is then determined from a continuum radiative
transfer calculation, in which the continuum opacity is provided by the dust.
It has been long debated whether this assumption still holds in the surface
layers of the disk, where the dust infrared emission features are produced.
In this paper we compute the temperature of the gas in the surface layers of
the disk in a self-consistent manner. The gas temperature is determined from
a heating-cooling balance equation in which processes such as photoelectric
heating, dissociative heating, dust-gas thermal heat exchange and line
cooling are included. The abundances of the dominant cooling species such as
CO, C, C$^+$ and O are determined from a chemical network based on the
atomic species H, He, C, O, S, Mg, Si, Fe 
\citep{kampbertoldi:2000}. The underlying disk models to our
calculations are the models of
\citet{dulvzadnat:2002}. We find that in general the dust and gas
temperature are equal to withing 10\% for A$_V\gtrsim$ 0.1, which
is above the location of the `super-heated surface layer' in which the dust
emission features are produced \citep[e.g.]{chianggold:1997}. 
High above the disk surface the gas 
temperature exceeds the dust temperature and can can become ---
in the presence of polycyclic aromatic hydrocarbons --- as high as 
600\,K at a radius of 100 AU.  This is a region
where CO has fully dissociated, but a significant fraction of hydrogen 
is still in molecular form. The
densities are still high enough for non-negligible H$_2$ emission to be
produced. At radii inward of 50
AU, the temperature of the gas above the photosphere can reach up to 
$\sim 10^4$ K. In the disk surface layers, the gas temperature 
exceeds the virial temperature of hydrogen. Some of this 
material could possibly evaporate, but firm conclusions have to await 
the fully self-consistent disk models, where the disk structure and gas 
temperature determination will be solved iteratively.
\end{abstract}


\section{Introduction}
The dusty gas disks surrounding T Tauri stars and Herbig Ae/Be stars are
believed to be the birthplaces of planets and have for this reason be the
subject of intense study in the last two decades. With the advent of high
sensitivity space infrared telescopes such as the Spitzer Space Telescope
(SST formally known as SIRTF), and ground-based infrared interferometers
such as the Very Large Telescope Interferometer (VLTI), these
`protoplanetary disks' can now be studied in unprecedented detail. This puts
a high pressure on theoretical modeling efforts of the structure and
appearance of these disks \citep[e.g.]{dalessiocanto:1998,chianggold:1997,
bellcassklhen:1997,lachmalbmon:2003,duldom2d:2004a}. 
Some of the simplifying assumptions that were
used so far may no longer be justified, since observations now start to be
able to distinguish between simplified and more realistic disk models. One
of the main assumptions underlying virtually all disk structure models to
date is the assumption of perfect thermal coupling between the dust and the
gas. Since the dust carries all the continuum opacity, it is the dust that
influences the continuum radiative transfer through the disk, and thereby
determines the dust temperature at every location in the disk. The
assumption that the gas temperature is equal to the dust temperature then
makes it possible to solve the equation of hydrostatic equilibrium, thus
yielding the density structure of the disk. The combined temperature and
density stucture of the disk constitutes the self-consistent disk model
from which the spectral energy disribution (SED) and images can be derived
and compared to observations.

However, it has long been debated whether the assumption of equal dust and
gas temperature holds in the surface layers of the disk 
\citep[e.g.]{chianggold:1997}. Above a certain height above the
midplane this dust-gas coupling is so low that the gas will find its own
thermal balance: a balance between gas heating processes such as the
photoelectric effect and dissociation processes on the one hand, and gas
cooling via atomic and molecular lines on the other hand. The structure of
this very tenuous upper part of the disk will then be similar to that of
PDRs, Photon-Dominated Regions \citep[e.g.]{th85,takahasietal:1991,SD95} 
as already noted by e.g.\ \citet{WL00} and \citet{vZadeletal:2003}. It is the
question whether this decoupling height is above or below the superheated
layer. It is also important to compute how the gas temperature will behave
above the decoupling height, and what consequences this has for observations
of certain molecular line species. 

It is the purpose of this paper to present 1+1D models for the vertical
temperature structure of the disk in which dust-gas decoupling is taken into
account. We first present the underlying vertical density structure
models and the physics that go into the self-consistent gas temperature
and chemistry calculation (Sect.~\ref{modeloverview}). Sect.~3 describes
the standard model including polycyclic aromatic hydrocarbons (PAHs) as 
well as a comparison model without
PAHs. The resulting gas temperatures and the chemical sturcture of these
models are presented in Sect.~3.1 and 3.2. In Sect.~4, we discuss the
possibility of disk evaporation as a consequence of the high gas temperatures 
in the upper layers of the disk. Sect.~5 summarizes our results and speculates
on future modeling and possible dynamical consequences.

\section{The model}
\label{modeloverview}
\begin{figure}[t]
\includegraphics[width=9cm]{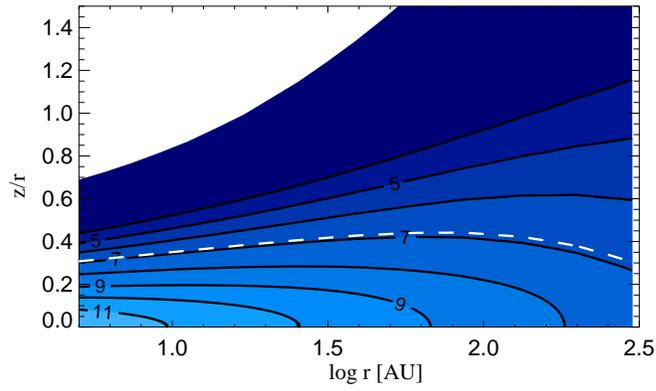}
\caption{\label{fig-dnz02-slice}
Vertical structure profile of the DZN02 disk model, which forms the
basis of the models described in the present paper. The contours 
denote the particle density $\log n$ in steps of 1. The white
dashed line shows the contour which includes 95\% of the disk
mass.}
\end{figure}

Our model calculations start from the 1+1D vertical density structure models
of \citet[henceforth DZN02]{dulvzadnat:2002}. 
In these models the dust and gas temperature were assumed to be
equal. In Fig.~\ref{fig-dnz02-slice} the density distribution of the model
is shown. 95\% of the mass reside inside the white dashed line. The left panel
of Fig.~\ref{tdusttgas} illustrates the dust temperature in this disk model.
As one can see, the dust temperature is lowest at the midplane.
The dust at the midplane is only heated by thermal infrared emission from
the surface layers. As one goes toward higher $z$, initially the dust
temperature rises slowly, but upon approaching the $\tau=1$ surface ($H_s$)
of the disk it rises more steeply. This is the transition from the optically
thick interior to the photosphere of the disk indicated by the dashed line
in Fig.~\ref{tdusttgas}). At even higher $z$ the dust
temperature profile levels off to a constant. This is the optically thin
dust temperature: these grains here are directly exposed to the central star
since they reside well above the photosphere. In the computation of this
model it was assumed that the gas temperature is equal to the dust
temperature, and therefore that the dust temperature distribution from
Fig.~\ref{tdusttgas} also represent the gas temperature distribution,
which enter into the equation of hydrostatic equilibrium. The dust grains
are 0.1~$\mu$m astronomical silicate grains (see DZN02 for more details).

\clearpage

\clearpage

In the present paper, however, we take this model one step further and
calculate the gas temperature self-consistently. However, we keep the
density structure and the {\em dust} temperature the same as the DZN02
models. In order to compute the gas temperature we need to solve the energy
balance equation for the gas.

\subsection{Chemistry}
In order to compute the abundances of the dominant cooling species, we must
include a chemical network into our model. The chemistry has been described
in detail in \citet{kampbertoldi:2000}, and is partly
based on rates from the UMIST database \citep{leteuff:2000}. The chemical
network consists of 47 species: H, H$^+$, H$^-$, H$_2$, H$_2^+$, H$_3^+$, He, 
He$^+$, C, C$^+$, O, O$^+$, S, S$^+$, Si, Si$^+$, Mg, Mg$^+$, Fe, Fe$^+$, CH, 
CH$^+$, CH$_2$, CH$_2^+$, CH$_3$, CH$_3^+$, CH$_4$, CH$_4^+$, CH$_5^+$, CO, 
CO$^+$, HCO, HCO$^+$, O$_2$, O$_2^+$, OH, OH$^+$, H$_2$O, H$_2$O$^+$, H$_3$O$^+$,
SiO, SiO$^+$, SiH, SiH$^+$, SiH$_2^+$, SiOH$^+$, e$^-$. Those species are
connected through 266 reactions including neutral-neutral, ion-molecule,
photoionization and photodissociation reactions. We account also for cosmic
ray induced photoreactions and charge exchange reactions. Neither grain surface
reactions nor ice formation are included, because the temperatures in the 
upper disk layers do not allow a significant freeze out of molecules. 

Some reactions are treated in more detail like the H$_2$ and CO photodissociation.
These two reaction rates are derived by using the UV flux and proper shielding
factors at the center of the dissociating band. For details we refer to 
\citet{kampbertoldi:2000}. C ionization is treated in a similar way by
integrating the photoionization cross section shortward of 1100~\AA\  and
accounting for H$_2$ and C self shielding. Depending on the He abundance and
on the way the C ionization is treated, the C$^+$/C/CO transition can shift 
as a function of depth in the disk: the relatively small fraction of He$^+$
obtained by cosmic ray ionization can be sufficient to destroy CO even at
locations, where it is already shielded from the stellar UV radiation field.
Table~\ref{abus} summarizes the element abundances used in the following 
calculations.
\begin{table}
\caption{Elemental abundances $\log \epsilon$ with respect to hydrogen in 
         the standard disk model.}
\begin{tabular}{llll}
element  &  $\log \epsilon$ \hspace*{1cm} & element  &  $\log \epsilon$ \\[2mm]
\hline\\
He       &  -1.125 & Si   & -6.100 \\
C        &  -3.879 & Mg   & -5.291 \\
O        &  -3.536 & Fe   & -5.208 \\
S        &  -5.728 &      &        \\
\end{tabular}
\label{abus}
\end{table}
The H$_2$ formation rate is given by
\begin{equation}
R_{\rm form} = 3\,10^{-17} \epsilon_{\rm H_2} \sqrt{\frac{T_{\rm gas}}{100}} \,\,\, ,
\end{equation}
where the new H$_2$ formation efficiency $\epsilon_{\rm H_2}$ of \citet{CT:02} 
was implemented to account for the reduced H$_2$ formation at high temperatures. 

The chemistry and gas temperature calculations are restricted to a single
dust particle size, in this case $a=0.1\mu$m, and a UV absorption cross
section of $\sigma_{\rm UV}=5.856\,10^{-22}$~cm$^2$/(H-atom), consistent 
with the optical properties of the DZN02 model.

We try to find stationary solutions to the
chemisty equations by linearizing the equations and solving a matrix
equation using an LU-decomposition method. It is not ensured that the
assumption of chemical equilibrium is justified. A time-dependent chemistry
may be a necessary next step. But since the simultaneous solution of the
time-dependent rate equations and the heating/cooling balance over the
entire life-time of the disk is quite time-consuming, we decided to use the
simplifying assumption of equilibrium photo-chemistry. Another simplifying
assumption which we put into our models is to assume that turbulent mixing
does not play a very important role. 

A disadvantage of focusing only on equilibrium solutions is that deep in the
optically thick regions of the disk no equilibrium solutions are
found. Whether this means that we are still missing physical ingredients in
our chemical network, or whether in reality the chemistry in the disk is
ever evolving, is not yet clear. But down to A$_V=20$ we can find stationary
solutions without a problem.

\subsection{Thermal balance of the gas}
The gas temperature is derived from balancing all important heating and
cooling processes:
\begin{equation}
\sum \Gamma (T_{\rm gas}, n_{\rm tot}, \epsilon_i, \chi) = 
\sum \Lambda (T_{\rm gas}, n_{\rm tot}, \epsilon_i)\,\,\,,
\end{equation}
where $\epsilon_i$ denote the abundances of the chemical species and $\chi$
the UV radiation field. As heating processes we include photoelectric
heating of dust grains, photodissociation of molecules and ionisation of
atoms, and cosmic rays. Cooling processes are molecular and atomic line
emission from species such as CO, H$_2$, CH, C, C$^+$, and O. This line
cooling is computed using an escape probability recipe, which we will
elaborate on below. Finally we include the thermal coupling between gas and
dust, which can be either a heating or a cooling term, dependent on the sign
of the temperature difference between gas and dust. Many of the details of
these processes are described in
\citet{kampzadel:2001}. But for the work presented here we added a
couple of features that were not described in that paper. They will be
described in the following paragraphs.

Some T Tauri disks are known to contain PAHs and small particles
\citep{Guertler:99}, but in a large fraction
of T Tauri star mid-IR spectra, the PAH features are not seen.
\citet{NattaKruegel:95} pointed out that this
might be due to the small beam sizes picking up only part of the
PAH emission and swamping by the disk continuum. Our standard
model uses the combined photoelectric heating rate for PAHs and small
graphitic particles by \citet{BakesTielens:94}.
In order to estimate the effect of PAHs on the gas
temperature in the upper layers of the disk, we also run a
model without PAH heating, using the heating rate for small interstellar
medium (ISM) particles by \citet{th85}. 
If photoelectric heating is the dominant mechanism, the resulting 
gas temperatures will be higher for the PAH+ISM mixture, than for 
the normal ISM dust. This is due to the higher yield of electrons
in the PAH+ISM mixture. Since PAHs do not contribute significantly 
to the total grain opacity, their influence on the underlying disk 
structure models is negligible. 

In addition, we implemented statistical equilibrium and infrared
pumping for the [C\,{\sc ii}] fine structure line cooling following
the same recipe as for O\,{\sc i} and CO in
\citet{kampzadel:2001}. To expand the line cooling to optically
thick lines, we use an escape probability mechanism, which is
discussed in the next paragraph.

\clearpage

\begin{table}
\caption{List of all heating and cooling processes that are used for
         the gas thermal balance}
\begin{tabular}{ll}
 heating                              & cooling \\[2mm]
\hline\\
photoelectric heating by PAHs         & Ly$\alpha$ line \\
and small carbonaceous grains         & \\[2mm]
collisional de-excitation of H$_2$    & O 6300 \AA\ line \\[1mm]
dissociative heating of H$_2$         & O fine structure lines \\[1mm]
formation heating of H$_2$            & C fine structure lines \\[1mm]
gas-grain collisions                  & C$^+$ 158~$\mu$m line \\[1mm]
C ionisation                          & H$_2$ ro-vibrational lines \\[1mm]
cosmic rays                           & CO rotational lines\\[1mm]
                                      & CH rotational lines\\[1mm]
                                      & gas-grain collisions \\
\end{tabular}
\label{tab:heatcool}
\end{table}

\clearpage

In Table~\ref{tab:heatcool}, we list all the processes that are included 
in the present model. Since the heating and cooling rates
depend on the abundances of the individual species, the chemistry
and the thermal balance have to be solved interatively. This is
done using Ridder's method and reevaluating the chemistry for
every new temperature guess. Convergence is typically achieved
within 10 steps. 

\subsection{Escape probability formalism}
The optical depth of molecular and atomic lines can be much higher than that
of the dust continuum. Therefore, high above the actual photosphere of the
disk the molecular lines can still be optically thick. Line cooling, even
above the continuum photosphere, is therefore a process in which radiative
transfer effects in the molecular lines must in principle be taken into
account. Unfortunately it is still quite challenging to compute the detailed
radiative transfer in all the molecular line species simultaneously with the
solution of a chemical network. As a first step we therefore employ an
escape probability formalism. Given a certain line optical depth $\tau$, the 
escape probability $\beta(\tau)$ describes the probability that a photon 
escapes from this line. The escape probability function is actually derived 
by an integration over all possible escape paths (inward directed photons are 
absorbed in the disk) and a gaussian line profile. This is a quite standard 
procedure in studies of PDRs, and therefore we adopt it here as well.

We use the formulas for $\beta(\tau)$ and line optical depth presented 
in the appendix B of \citet{th85}. The optical depth in all cooling
lines in vertical direction $z$ is
\begin{equation}
\tau (z) = A_{ul} \frac{c^3}{8 \pi \nu^3 \delta v} \int_0^z n_u(z') 
          \left(\frac{n_l(z') g_u}{n_u(z') g_l} -1 \right) dz'\,\,\,.
\end{equation}
$A_{ul}$ is the transition probability of the line, $c$ the velocity of
light, $\nu$ the line frequency, $\delta v$ the line broadening, $n_u$
and $n_l$ the upper and lower level densities, and $g_u$ and $g_l$ the
upper and lower level statistical weights. The escape probability
can be derived from the optical depth $\tau$ 
\begin{eqnarray}
\beta(\tau) & = & \frac{1 - \exp{-2.34 \tau}}{4.68 \tau}\,\, , \hspace*{15mm} \tau < 7 \nonumber \\
            & = & \left( 4\tau \left( \ln{\frac{\tau}{\sqrt{\pi}}} \right)^{0.5} \right)^{-1}
                  \,\, , \hspace*{5mm} \tau \geq 7 \\
\end{eqnarray}
The maximum escape probability 
is 0.5, because of the slab geometry.

\subsection{Flaring angle recipe}
The physics of the upper layers of the disk is, similar to PDRs, 
regulated by the influx of UV and optical radiation from an
external source. In our case this is the central star. In ordinary PDR
calculations the source of UV radiation is assumed to illuminate the surface
of the PDR from above (under a 90 degree angle from above). In contrast, for
protoplanetary disks, the light of the central star penetrates the surface
of the disk under a very shallow grazing angle (called the `flaring angle'),
typically of the order of $\alpha=0.05$. The definition of the flaring
angle is:
\begin{equation}
\alpha \equiv R\frac{d}{dR}\left(\frac{H_s}{R}\right)
\end{equation}
where $H_s$ is the height of the continuum $\tau_V=1$ surface for photons
emerging from the central star. A very rough estimate of $H_s$ is the height
$z$ above the midplane where the temperature of the dust is
$\exp(-1/4)=0.7788$ times the optically thin dust temperature (see
dashed line in Fig.~\ref{tdusttgas}).

Because the grazing incident angle $\alpha$ is usually very small, stellar
radiation can penetrate only very limitedly into the disk, unless dust
scattering deflects the UV photons deeper into the disk \citep{vZadeletal:2003}.
The way in which this grazing
incident angle $\alpha$ is included in the model is as follows. We assume,
first of all, that there is no scattering which could deflect ionizing
radiation deep into the disk's photosphere. We can then simply `mimic' the
effect of the longer path (compared to vertical incidence) by multiplying
all extinction effects of the stellar radiation by $1/\alpha$. The reason is
that for each cm in vertical direction, the photons pass in reality through
$1/\alpha$ cm of gas by virtue of the grazing angle of $\alpha$. This means
that effectively (per cm in vertical direction) the (self-)shielding of
molecular species is $1/\alpha$ times so strong as when the UV source
illuminated the layers perfectly from above.  Also the radiation responsible
for the photoelectric heating effect are extincted by a factor of $1/\alpha$
more efficiently compared to normal PDRs. And finally, a similar effect
takes place for the radiation that excites the higher levels of cooling
species. All in all
this has the effect that much of the `action' takes place at lower A$_V$
than would be the case for normal PDRs. For instance, dissociation
boundaries lie higher up in the atmosphere of the disk, and dust-gas
decoupling may take place at lower A$_V$ than would be the case if the UV
source would shine 90 degrees from above. This recipe of enhancing the
extinction of stellar light by a factor $1/\alpha $ in order to mimic the
effect of shallow incidence is called the `flaring angle recipe'. For the
cooling via molecular lines such a flaring angle recipe is not used, because
the emitted photons can escape in all directions.

\subsection{Grid resolution and numerical procedure}

It is of essential importance to resolve the disk in vertical direction in 
a way that is appropriate for the disk chemistry. Initially, the 1+1D vertical 
density structure model grid is adapted to resolve the dust continuum optical 
depth scale. However, the gas chemistry takes place on a different scale. To ensure
that we resolve e.g.\ the H/H$_2$ transition, we resampled the grid to a ten 
times finer resolution by linear interpolation.

The calculation of the combined heating/cooling balance and chemistry is
extremely time consuming and should be restricted to those regions, where
T$_{\rm gas} \neq$ T$_{\rm dust}$. However, a simple check at a single grid
point is not sufficient to determine, whether T$_{\rm gas}$ has converged
to T$_{\rm dust}$, because it could actually be a crossing point. We implemented 
here a criterium that checks for a gradual approach of T$_{\rm gas}$ and 
T$_{\rm dust}$: If both temperatures agree to within 2\% over an interval 
$\delta z/z \sim 0.3$, the detailed balance calculations are stopped and 
T$_{\rm gas} =$ T$_{\rm dust}$ is assumed below this point.

\section{Results}
We present here one standard model, and some features from a model without
PAHs. Our standard parameter set is as follows. For the central star
we take a T Tauri star with $M_{*}=0.5\,M_{\odot}$, $R_{*}=2.5\,R_{\odot}$
and $T_{*}=4000\,$K where $T_{*}$ is the effective temperature of the
photosphere of the star. Our disk model is assumed to have a surface density
profile $\Sigma(R)$ given by a powerlaw:
\begin{equation}
\Sigma(R) = \Sigma_0\;\left(\frac{R}{\rm AU}\right)^{-1}
\end{equation}
with $\Sigma_0=50$~g~cm$^{-2}$. The inner radius of the disk
is taken to be $R_{\rm in}=0.1$~AU, and the outer radius
$R_{\rm out}=300$~AU.  With these parameters the mass of the disk
becomes $M_{\rm disk}= 0.01~M_{\odot}$. The flaring angle for this
model follows automatically from the solution to the equations of DZN02, and
we use this flaring angle (which is a function of $R$) as input to the gas
heating/cooling and chemistry calculations.

\clearpage

\begin{figure}[h]
\includegraphics[width=9cm]{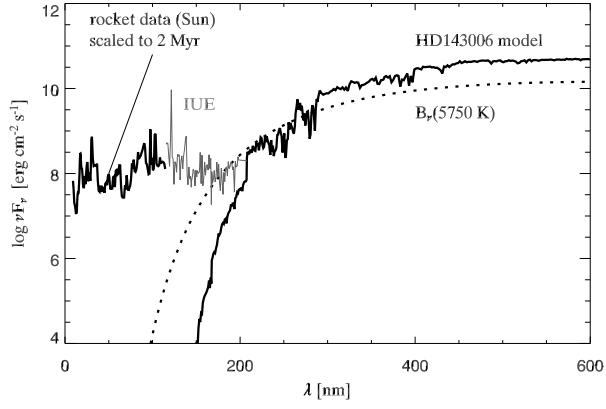}
\caption{\label{radfield}
Radiation field of a 2~Myr old T Tauri star: spectrum composed
from scaled solar UV observations, IUE data of HD\,143006 and a
solar-type Kurucz model.}
\end{figure}

\clearpage

In order to simulate the surrounding remnant medium of the star formation
region, we extend the disk model to larger heights assuming a smooth
transition to a constant low density region of $n=5000$~cm$^{-3}$; this is
a typical value for the remnant molecular cloud gas as seen in the low-mass
star forming regions Taurus Aurigae or Chamaeleon \citep{pallastahler:2002,
mizunoetal:2001}.
The radiation field is assumed to be that of an active young T Tauri
star. Following the recipe described in \citet{kampsammar:2003}, 
we scaled the solar chromospheric activity
backwards to an age of 2\,Myr, assuming an age of 4.6~Gyr for our Sun. 
The integrated stellar UV flux is then calculated to be
\begin{equation}
\chi= \int_{912\,{\rm \AA}}^{1110\,{\rm \AA}} \frac{1}{h\nu} F_\nu d\nu = 
  2.63\,10^{18}~{\rm cm^{-2}~s^{-1}}
\end{equation}
at the stellar surface. This value is diluted with $(R/R_{*})^2$ according to
the respective distance of the slab to the star. It corresponds to 
$1.7\,10^{5}$ times the interstellar Draine field at 10 AU. Fig.~\ref{radfield} 
shows the complete stellar spectrum assumed for the T Tauri star. For the composed
spectrum, we used the scaled solar UV flux, the IUE observations of the T
Tauri star HD\,143006 and a Kurucz model with $T_{\rm eff} = 5750$\,K, $\log
g = 4.5$.

Since we do not assume any turbulent mixing, the disk model is divided into
vertical slabs and a 1D coupled chemical and heating/cooling calculation is
carried out starting at the upper
boundary of the model.  Due to the strong radiation field impinging on the
inner disk, which complicates the convergence of the model, we only
start our model calculations at 5~AU.

\clearpage

\begin{figure*}[t]
\includegraphics[width=18cm]{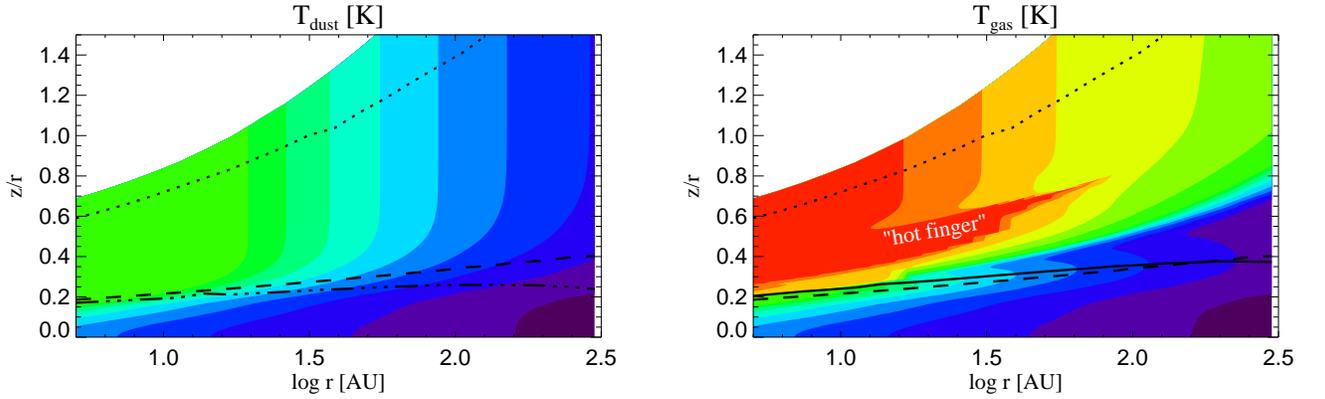}
\caption{\label{tdusttgas}
Dust and gas temperature in the standard disk model. The solid
line indicates the location, below which gas and dust temperatures agree to within 10\%
(right panel). If over an interval $\delta z/z \sim 0.3$ both temperatures agree within 2\%,
complete coupling, that is T$_{\rm gas} =$ T$_{\rm dust}$, is adopted below this depth 
(dashed-dotted line in left panel). The 10\% agreement occurs at or slightly above the 
$\tau = 1$ surface layer (dashed line). The region around the $\tau = 1$ surface, where 
the gas temperature contours change their slope and the gas temperature drops below the 
dust temperature, is referred to as the ``undershoot region'' (see Fig.~\ref{20AUslice}
and \ref{100AUslice}. The dotted line marks the transition between the surrounding cloud and
the disk. The colors correspond to the following temperatures: 10, 20, 30, 40, 50, 60, 
70, 80, 90, 100, 200, 500, 1000, 2000, and 5000\,K.}
\end{figure*}

\begin{figure}[h]
\includegraphics[width=9cm]{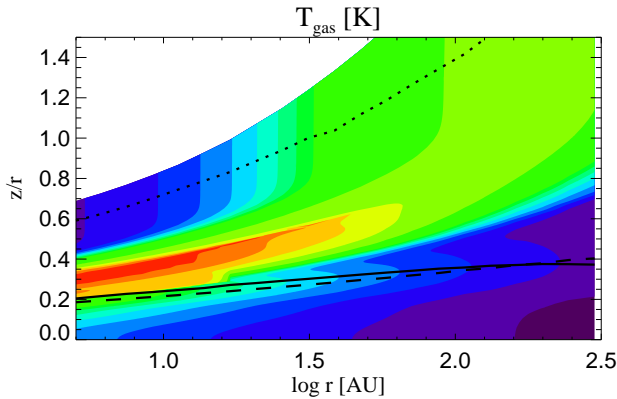}
\caption{\label{tdusttgas_noPAH}
Gas temperature in the disk model without PAHs. See caption of
Fig.~\ref{tdusttgas} for the meaning of the lines and colours.} 
\end{figure}

\clearpage

\subsection{Gas temperature}

\begin{figure*}
\includegraphics[width=18cm]{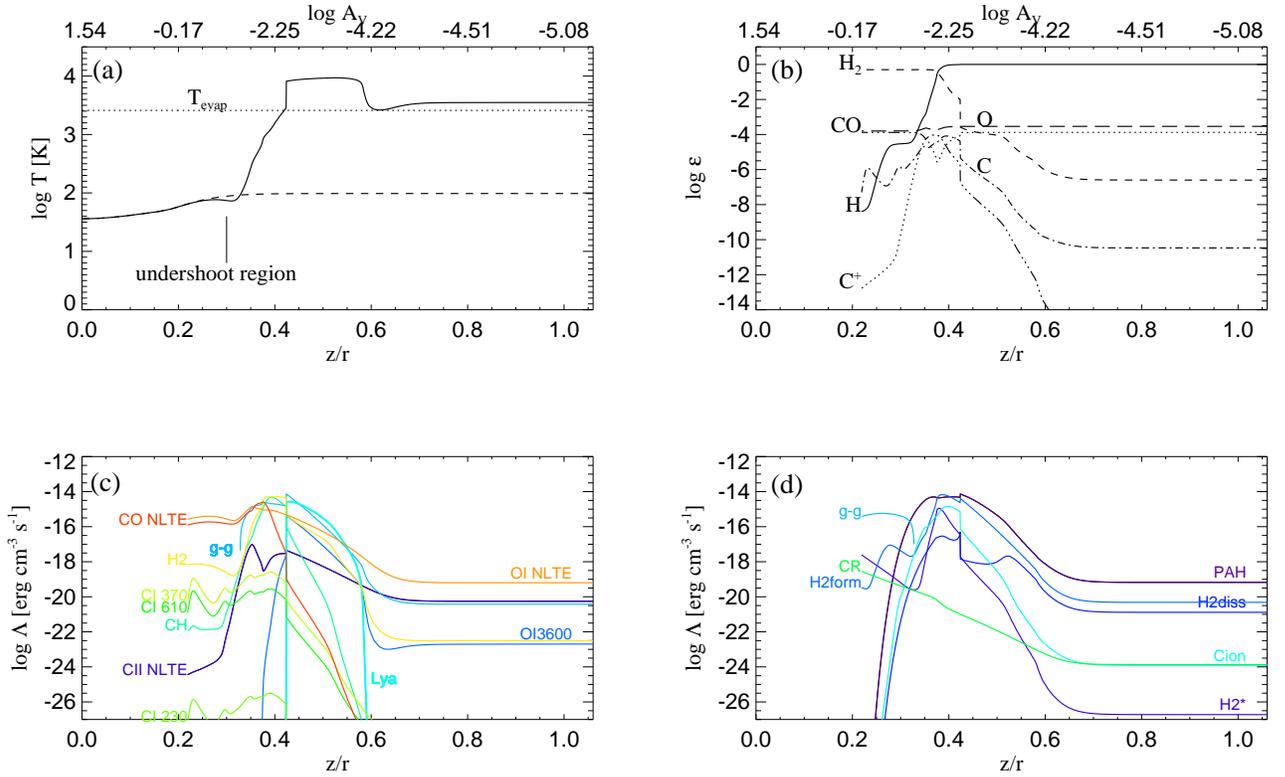}
\caption{\label{20AUslice}
A slice through the disk model at 20\,AU. (a) gas temperature (solid line),
dust temperature (dashed line) and evaporation temperature of an H-atom.
(b) abundances $\epsilon$ of important chemical species, H$_2$ (solid line), H
(dashed line), C$^+$ (dotted line), C (dash-dotted line), CO (dash-three-dotted
line), and O (long-dashed line). The upper scale in these two panels illustrates
the vertical visual extinction into the disk at selective points. Panels (c) 
and (d) denote the individual cooling and heating processes as a function of 
disk height $z/r$.}
\end{figure*}

\begin{figure*}
\includegraphics[width=18cm]{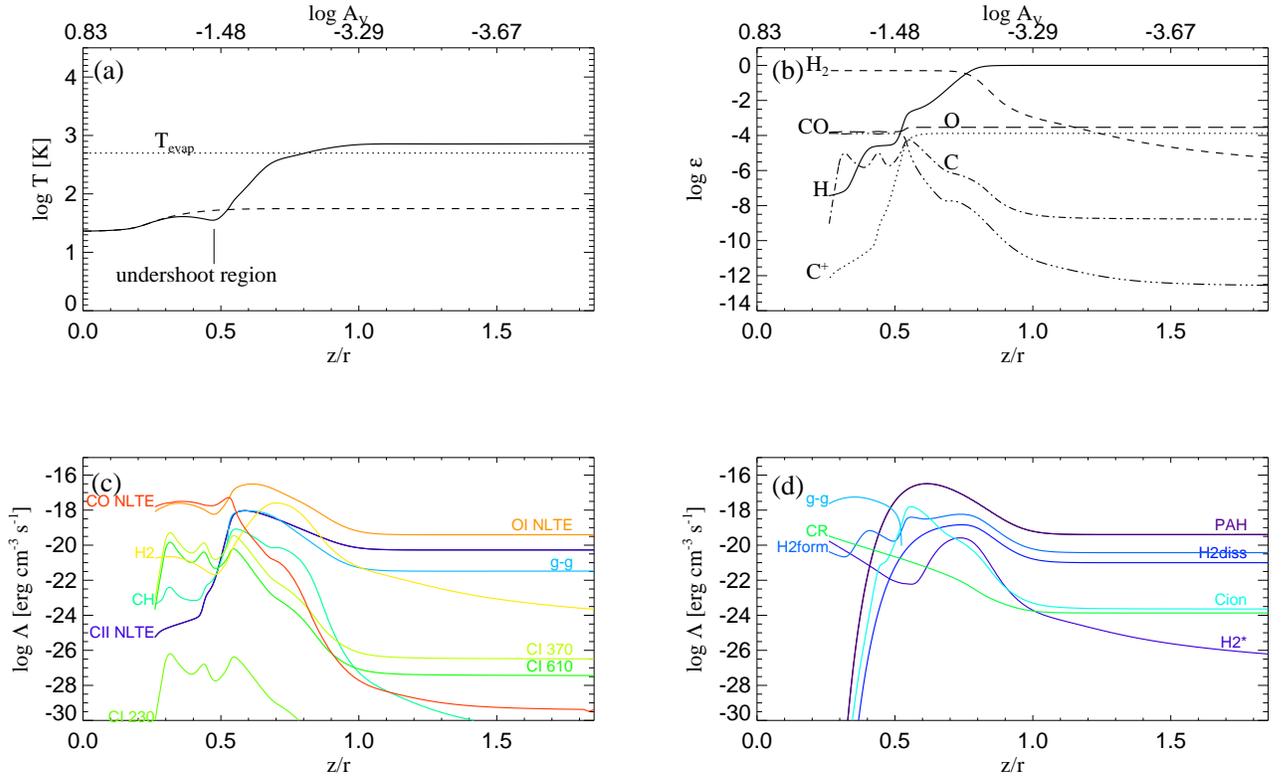}
\caption{\label{100AUslice}
A slice through the disk model at 100\,AU. See Fig.~\ref{20AUslice} for
the meaning of the lines.}
\end{figure*}

The resulting gas temperature for the standard model is shown in 
Fig.~\ref{tdusttgas} together with the dust temperature from the input model. 
The gas temperature in the disk atmosphere is much larger than the dust
temperature. Deeper in the disk, slightly above the $\tau = 1$ surface, 
gas and dust are well collisionally coupled and hence in thermal
equilibrium (within a 10\% error margin). A very pronounced feature 
of these models is the hot region,
$T_{\rm gas} \approx 10\,000$~K in the disk atmosphere at radii smaller than
50~AU, which we will address as ``hot finger'' in the following (the region 
is labelled in the right panel of Fig.~\ref{tdusttgas}). Another 
feature of these models is the ``undershoot region'' for the gas temperature 
around the $\tau = 1$ surface. This is the region, where the gas temperature
contours in Fig.~\ref{tdusttgas} change their slope and the gas 
temperature drops actually below
the dust temperature. For a more quantitative illustration of this model, 
we included two perpendicular slices at 20 and 100~AU (Fig.~\ref{20AUslice} 
and \ref{100AUslice}), where we indicate the location of the ``undershoot
region''.

The second model without PAH heating is shown in Fig.~\ref{tdusttgas_noPAH}.
The lower equilibrium gas temperatures in the surface layers are due to the lack
of PAH heating as compared to the standard model. Nevertheless, it is important
to notice that the lower part of the ``hot finger'' is still present in this
model. As for the standard model, the assumption of T$_{\rm gas} \sim$ 
T$_{\rm dust}$ is satisfied below the $\tau = 1$ surface.

All the above described features are explained in the following paragraphs 
by a complex combination between gas chemistry and gas heating/cooling processes. 
We will concentrate on the standard model and explain only differences arising 
from the absence of PAHs in the second model.

\clearpage

\clearpage

The heating is dominated by photoelectric heating (PE) due to PAHs and small
carbonaceous dust grains. At the lower boundary of the ``hot finger'',
where the electron abundance and hence the PE heating drops, the gas is 
heated primarily by H$_2$ formation (see Fig.~\ref{20AUslice}d). As H$_2$ 
becomes the dominant carrier of hydrogen deeper in the disk, this heating 
process turns off. We verified that our models resolve the H/H$_2$ transition.
For a more detailed discussion of the H$_2$ chemistry, we refer to the next paragraph.

Cooling is mainly due to [O\,{\sc i}] fine structure lines. 
In the ``hot finger'' region, Lyman\,$\alpha$ and H$_2$ ro-vibrational
line cooling are the dominant cooling processes followed by gas-grain 
collisions (see Fig.~\ref{20AUslice}c). At larger radii, beyond 
the ``hot finger'' region, the balance between [O\,{\sc i}] fine 
structure cooling and PE heating sets the gas temperature 
(Fig.~\ref{100AUslice}c and d).

The steep temperature rise at the upper boundary of the hot finger is due
to a different scaling behaviour of the main heating and cooling rates,
photoelectric heating versus [O\,{\sc i}] line cooling. PE heating scales 
roughly as $n^{1.8}$: it depends linearly on the grain number density,
which is linked by a constant to the total number density, and it depends
indirectly on electron density $n_{\rm e}^{0.8}$ through the PE heating 
efficiency. The higher the electron density, the lower the grain charge
and hence the larger the efficiency of the PE heating. Given the strong 
radiation fields that illuminate the surface of the disk, carbon is the 
main source of electrons, because it is fully ionized: $n_{\rm e} 
\sim n_{\rm C^+} \sim n \cdot \epsilon_{\rm C}$. While the PE heating scales 
with $n^{1.8}$, cooling is dominated by [O\,{\sc i}] fine structure lines 
in NLTE and hence scales with $n$. If the density rises, 
the increase in PE heating is much larger than the increase in cooling. 
Therefore, the gas temperature rises.
Since PE heating depends only weakly on temperature $T^{-0.34}$, 
$T_{\rm gas}$ has to increase strongly to compensate the density effect.
The temperature levels off again as LTE is reached and the cooling lines 
are collisionally dominated and scaling with $n^2$. The high 
temperatures obtained in this model depend sensitively on the atomic
data that enters the calculation of the cooling lines. The reason
is that the PE heating depends only weakly on the gas temperature
in the regime of high irradiation. Therefore, any small change in 
the cooling rates --- due to e.g.\ uncertainties in the collision
cross sections --- causes a large change in the equilibrium solution
of the gas temperature. However, this holds only for the uppermost
layers of the disk. In the ``hot finger'' region, densities are
typical of the order of $0.5-1\times 10^6$~cm$^{-3}$, and O\,{\sc i} level
population numbers deviate by less than 10\% from LTE. Thus, the 
gas temperature in this region is hardly affected by uncertainties in
the O-H or O-e$^-$ collision cross sections.

In the inner regions of the disk, $r<50$~AU, the gas temperature
shows a steep temperature gradient just below the hot finger 
(Fig.~\ref{20AUslice}a). This is due to the large drop in H$_2$ 
formation heating as most of the hydrogen is turned into molecular hydrogen.

The undershoot feature near the $\tau = 1$ surface is an intrinsic 
feature that is due to the physical description of the gas-grain
interaction. [O\,{\sc i}] and CO cooling dominate the energy balance 
in the regions, where T$_{\rm gas} \neq$ T$_{\rm dust}$. As the density rises
towards the midplane, the line cooling is so efficient that T$_{\rm gas}$ drops
even below T$_{\rm dust}$. This feature is not present in the model calculations
of \citet{jonkheid:2004}. If the gas-grain collisional coupling would be larger
or the [O\,{\sc i}] fine-structure line cooling smaller, gas and dust would
start to couple at higher $z$ in the disk and hence the undershoot region
would vanish. This could e.g.\ be achieved by a different density gradient 
([O\,{\sc i}] fine-structure lines become optically thick at higher $z$) 
and/or smaller grains (gas-grain collisional cooling is enhanced compared 
to [O\,{\sc i}] fine-structure lines and CO cooling). We assumed here, that 
the 0.1~$\mu$m grains are responsible for the gas-grain cooling, not the small 
PAHs.

In the model without PAHs, the lower part of the ``hot finger''
is still present (Fig.~\ref{tdusttgas_noPAH}). In this part, 
PE heating is not the dominant heating process and hence the omission of 
PAHs does not change the overall picture. Mainly, H$_2$ formation heating 
is balanced by H$_2$ ro-vibrational line cooling and gas-grain collisions. 
Hence, the resulting gas temperature is the same as in the standard
model. However, the upper layers of the disk without PAHs
are much cooler than in the standard model. [C$^+$] 158~$\mu$m line
cooling is now the dominant cooling process inwards of $\sim 60$~AU. 
Interestingly, the gas temperature drops as one gets closer to the star. 
This seems at first glance counterintuitive, because the strength of the 
radiation field increases and therefore one would expect a stronger 
PE heating rate
\begin{equation}
\Gamma_{\rm PE} \sim \epsilon_{\rm PE}\,\, \chi\, n_{\rm tot}\,\,\, .
\end{equation}
However, the PE heating efficiency $\epsilon_{\rm PE}$ drops in the absence 
of PAHs much faster than $\chi$ and thus the heating rate becomes smaller. 
The ISM grains, which are rather large compared to the PAHs, become strongly 
charged in the presence of high UV radiation fields. Therefore it becomes
more difficult for an electron to escape the grain and actually heat the 
gas. In the model with PAHs, the slope of the PE heating efficiency is
much flatter, because the PAHs, which are actually large molecules, are
less charged than the ISM grains.

\subsection{Gas chemistry}

\begin{figure*}
\includegraphics[width=16cm]{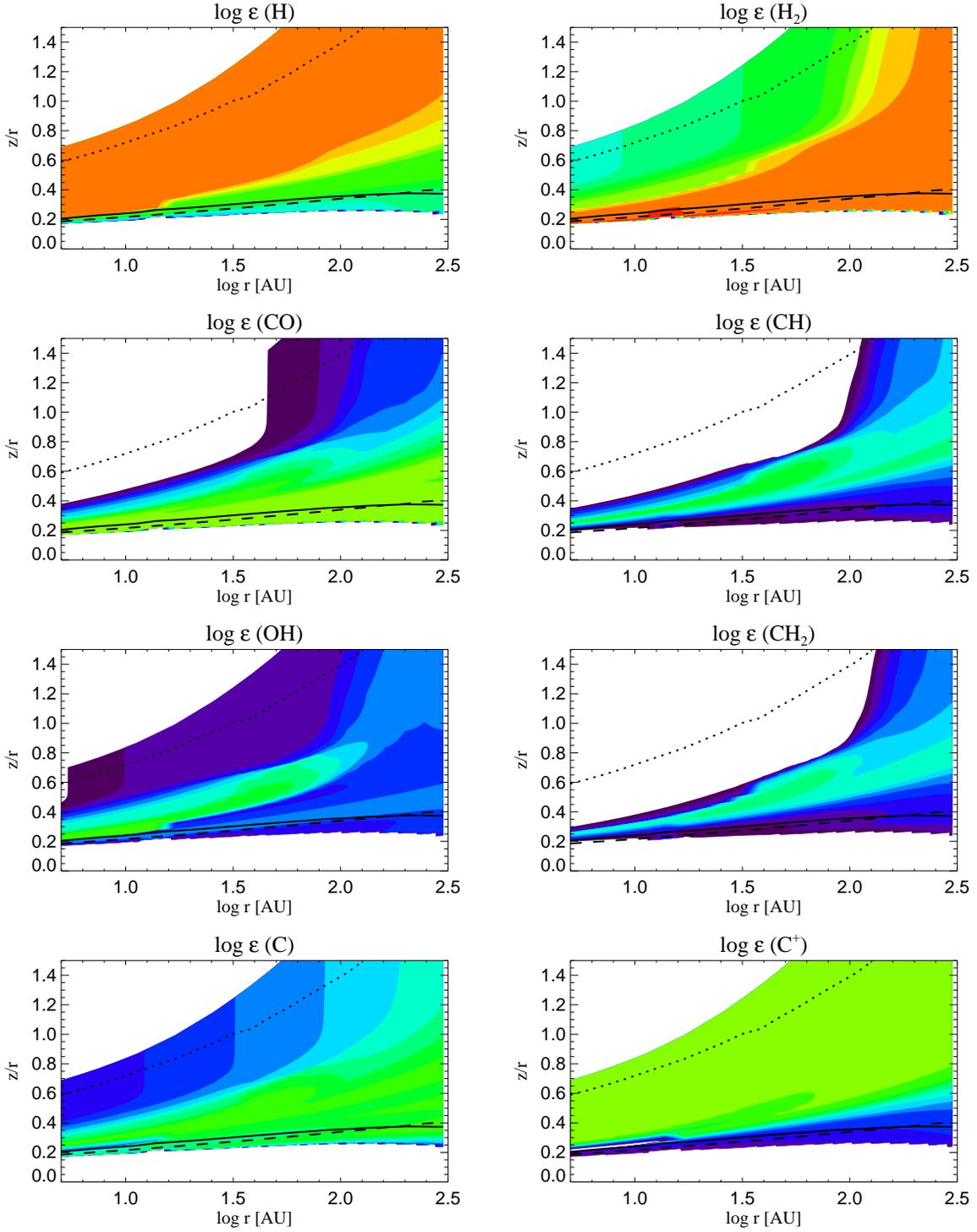}
\caption{\label{chem}
Abundances of selected species in the standard disk model: H, H$_2$, C,
C$^+$, CO, CH, OH, and CH$_2$. See Fig.~\ref{tdusttgas} for the meaning
of the overplotted lines. The colors correspond to logarithmic abundances
in steps of 1 and range from $\log \epsilon = -14$ to $0$.}
\end{figure*}

Fig.~\ref{chem} illustrates the abundances of the most important atoms and
molecules in the disk. The H/H$_2$ transition occurs well above the
warm disk surface layer. The same holds for the C$^+$/C/CO
transition. Interestingly, the hot finger is almost devoid of H$_2$, because
it is chemically destroyed at high temperatures. At low A$_{\rm V}$
($\log {\rm A}_{\rm V} < -2$), H$_2$ is photodissociated, but deeper in
the disk, the UV radiation is efficiently shielded and H$_2$ is mainly
destroyed via collisions with first H-atoms and then O atoms, leading to 
the formation of H and OH. The latter is subsequently photodissociated 
into its atomic components. Hence, H$_2$ is destroyed chemically and
not via direct photodissociation by stellar UV photons. There exists a 
region in our model, where the CO photodissociation front lies at the same 
A$_{\rm V}$ than the H$_2$ dissociation front (see Fig.~\ref{20AUslice}b).
Leaving out the PAHs does not change the resulting chemical
structure of the model.

An interesting feature is the ``hot finger'' region, which shows
a very high abundance of OH. Similarly, there are intermediate layers
in the disk atmosphere, where CH and CH$_2$ show an abundance maximum.
Although these layers do not contain a lot of disk mass, they may still
be observable. The column densities of CO, CH, and CH$_2$ are $3.9\,10^{18}$,
$2.8\,10^{13}$, and $6.6\,10^{12}$~cm$^{-2}$ at a distance of 100~AU 
for the standard model.

The disk atmosphere contains a significant amount of gas in molecular form. 
The high temperature of this gas might help to excite e.g.\ molecules such 
as H$_2$, which would otherwise not be observable.

\clearpage

\clearpage

\section{Disk evaporation}

\begin{figure}[t]
\includegraphics[width=9cm]{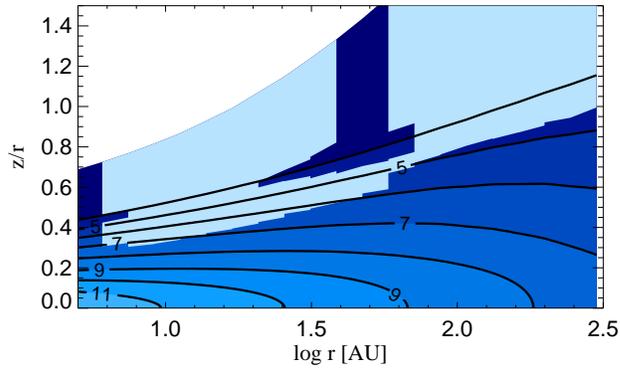}
\caption{\label{evap}
Same as Fig.~\ref{fig-dnz02-slice}, but now the light area indicates the 
the region, from which hydrogen gas can escape the standard disk model.}
\end{figure}

Given the gas temperatures derived for the disk model with PAHs, H\,{\sc i}
will evaporate from the upper disk layers. Its thermal velocity
$v_{\rm H}$ in the standard model is large enough to overcome gravity. However, 
it exceeds the escape velocity
\begin{equation}
v_{\rm esc} = \sqrt{\frac{2 G M_\ast}{r}}
\end{equation}
only marginally. $G$ is here the gravitational constant, $M_\ast$ the stellar 
mass and $r$ the distance from the star. Fig.~\ref{evap} shows the region, 
where hydrogen gas can escape given the rather conservative criterium
\begin{equation}
T_{\rm gas} > T_{\rm escape} = \frac{G M_\ast m_{\rm H}}{k_{\rm B} r} \,\,\, . 
\end{equation}
$m_{\rm H}$ is the mass of a hydrogen atom and $k_{\rm B}$ the Boltzmann
constant.

The thermal velocities $v_i$ of the heavier elements are smaller than the 
escape velocity and they could only escape if they are entrained with the
evaporating hydrogen flow. Given a collision rate $C(v_i)$, their 
mean free path $l$ in a hydrogen flow is \citep{Kwok:1975}
\begin{eqnarray}
l & = & \frac{v_i}{C(v_i)} \nonumber \\
  & = & \frac{1}{2\pi a_0^2}\,\, \sqrt{1+\frac{m_i}{m_{\rm H}}}\,\, n_{\rm H}^{-1} \nonumber \\
  & \approx & 379.9 \,\, \sqrt{1+\frac{m_i}{m_{\rm H}}}\,\, n_{\rm H}^{-1}~{\rm AU} \,\,\, . \\ 
\end{eqnarray}
$m_i$ is the masses of the heavy element, $a_0$ is the Bohr radius, and $n_{\rm H}$ 
is the hydrogen particle density. Assuming the minimum density from our models, 
5000~cm$^{-3}$ in the surrounding medium, we find that the maximum free path for 
an oxygen atom is $\sim 0.3$~AU. Although the mean free path of the heavy elements 
is rather small, the force acting on these particles is smaller than gravity. 

Firm conclusions on the possibility of evaporation can only be drawn from dynamical 
calculations, which account for additional forces like the pressure gradient
and centrifugal term. The deviation from Keplerian velocity is approximated as
\begin{eqnarray}
dV & \sim & \frac{c_s^2}{2 \Omega r} \frac{\delta ln \rho}{\delta ln r} \nonumber \\
   & =    & -5.93 \left( \frac{r_{\rm AU} M_\odot}{M_\ast} \right)^{0.5} T_{\rm gas} p \\
\end{eqnarray}
\citep{klahr:2000}, where $\Omega$ is the Keplerian frequency, $c_s$ the sound speed,
and $\rho \sim r^{-p}$ the gas mass density. Although $dV$ is rather small in the cold
disk midplane, the warm disk surface ($T_{\rm gas} \sim 1000$~K) will rotate
substantially subkeplerian, $dV/V_{\rm kep} \gtrsim 0.2$. This may also give rise 
to shear instability.

\clearpage

\clearpage

It should however be kept in mind, that there is some uncertainty in 
the gas temperature due to the atomic data that enters the cooling
line calculations and the rather simple fitting formulas used for the
heating processes. However, \citet{Adams:2004} have recently shown that the
evaporation flow starts indeed already at subsonic velocities. In our case,
this would mean that the evaporation would start deeper in the disk, leading to
a much higher mass loss rate. The uncertainty in the gas temperature will then
be less important, because rather than switching from non-evaporation to evaporation, 
the mass loss process would be rather smooth and differ only in its efficiency.
In the model without PAHs, the gas temperature is generally lower and disk evaporation 
is thus less effective.

\section{Conclusions}

We find that the dust and gas temperature in protoplanetary disks around T
Tauri stars are well coupled up to the location of the
`superheated surface layer' where the dust emission features are produced.
This means that for models aimed at predicting dust continuum
emission from protoplanetary disks, the assumption of equal gas and dust
temperature is probably not too far from the truth. 

At radii smaller than 50~AU, the gas temperature can reach values
of $\sim 10\,000$ K above the disk photosphere. First estimates
indicate that at least hydrogen can evaporate from the disk surface.
To verify this, an iteration procedure is required, in which the resulting gas 
temperature structure is inserted into the equation of hydrostatic 
equilibrium, yielding a new density structure, which then feeds back 
into the chemistry-heating-cooling model. This, however, is beyond 
the scope of the present paper. We can nevertheless speculate
on the consequences of such an iteration given the results
of our present model: in the upper layers of the innermost disk regions, 
$r<15$~AU, the gas temperature depends hardly on density. Therefore, we 
do not expect our gas temperature structure to change a lot in these 
regions, if the feedback on the disk structure would be taken
into account (more flaring in the outer parts and hence a different
density structure). But at larger radii, $r>15$~AU, the ``hot finger'' 
and hence the evaporating region, could vanish if the density structure
of the disk changes.

It is especially interesting to note the importance of PAHs on
the resulting gas temperature structure of the upper disk atmosphere.
PAHs are a significant heating source in the uppermost layers. The
presence of PAHs in the calculations presented here is necessary
for the possibility of disk evaporation. It does, however, not 
lead to significant changes in the chemical structure of the model.

Above the superheated surface layer, the dust and the gas thermally
decouple. Some molecular species, such as H$_2$, CO, CH, CH$_2$, OH, probe
the very upper layers of the disk, in the region where the gas temperature 
has already decoupled from the dust. In order to make reliable predictions 
for these molecular lines, a detailed gas temperature and molecular abundance 
calculation of the kind presented here is important. \citet{jonkheid:2004}
found strong increases, up to a factor 10, in the [O\,{\sc i}] and [C\,{\sc ii}] 
fine structure line fluxes, due to different excitation conditions in the
disk models with a detailed calculation of the gas temperature. The accuracy
of the gas temperature determination is extremely difficult to assess,
especially given the fact that atomic and molecular collision cross 
sections are often poorly known at high temperatures and our limited
knowledge of the grain composition and size in these disks. There are
e.g.\ only very few direct detections of PAHs in T Tauri disks. It
is very important for the models to seek for input from the observations. 
Instruments like VISIR at the VLT have the high resolution ($R\sim 
12\,500-25\,000$) that is necessary to pick up the narrow gas lines from 
these disks. If such gas line observations would reveal the presence of 
an extremely hot disk surface layer, we might have an indirect evidence 
for the presence of PAHs in disks around T Tauri stars.

In this paper we have always assumed that the dust and the gas are
well-mixed, i.e.~that the dust has not settled toward lower elevations above
the midplane. In reality, however, in the very surface layers of the disk,
this settling process takes place on a very short time scale. At 10 AU, for
instance, a dust grain of 0.1 $\mu$m settles below the initial photospheric
height $H_s$ in less than 10$^3$ years. A detailed study of the consequences that
this settling process has on the appearance of the disk has been given recently 
by \citep[in press]{duldom2d:2004b}. In future work this process will have 
to be included into the present model, in order to study the effect of dust 
settling on the temperature structure of the disk.

Finally, some short speculations on the dynamical consequences of our
gas temperature results: the strong temperature gradients, especially above
the hot finger, suggest that convective instabilities might occur. Material
expanding into the upper layers of the disk atmosphere will encounter a
steep negative temperature gradient that might exceed the adiabatic one.
In addition, these strong temperature gradients could also give rise to 
shear instabilities. Such convective and shear instabilities could lead to 
additional mixing, especially in the upper layers of the disk, where one 
would like to keep the small dust grains.

\acknowledgments

We are grateful to Ewine van Dishoeck and Bastiaan Jonkheid for a
detailed comparison of our codes and numerous discussions on the
details of thermal balance calculations. We acknowledge also the
fruitful discussions on this topic taking place during an ongoing 
detailed comparison of several PDR codes. Furthermore, we would like 
to thank David Hollenbach and Doug Lin for interesting discussions on 
the dynamical consequences of these models during a workshop on 
protoplanetary disks held at Ringberg castle. An anonymous referee
helped us to improve the perspicuity of the paper significantly. 
I.~Kamp acknowledges support by a grant from the Netherlands 
Organisation of Scientific Research (NWO).

\end{document}